\documentstyle[12pt]{article}
\def\eql{\stackrel{\rm law}{=}}
\newcommand{\be}{\begin{equation}}
\newcommand{\ee}{\end{equation}}
\newcommand{\ba}{\begin{eqnarray}}
\newcommand{\ea}{\end{eqnarray}}

\begin{document}
\setlength{\baselineskip}{.7cm}
\renewcommand{\thefootnote}{\fnsymbol{footnote}}
\sloppy

\begin{center}
{\Large \bf Convergent multiplicative processes repelled from zero: power
laws and truncated power laws}

\vskip .17in
Didier Sornette$^{1,2}$ and Rama Cont$^1$

{\it $^1$ Laboratoire de Physique de la Mati\`ere Condens\'ee, CNRS
URA190\\ Universit\'e des Sciences, B.P. 70, Parc Valrose, 06108 Nice Cedex
2,
France \\

$^2$ Department of Earth and Space Science\\ and Institute of Geophysics
and
Planetary Physics\\ University of California, Los Angeles, California
90095\\
E-mail : cont@ens.fr
}
\end{center}

\newpage

\begin{abstract}
Levy and Solomon have found that random multiplicative processes $w_t =
\lambda_1 \lambda_2 ... \lambda_t$  (with $\lambda_j > 0$) lead, in the
presence of a boundary constraint, to a distribution $P(w_t)$ in the form of
a power law $w_t^{-(1+\mu)}$. We provide a simple exact physically intuitive
derivation of this result based on a random walk analogy and show the
following: 1) the result applies to the asymptotic ($t \to \infty$)
distribution of $w_t$ and should be distinguished from the central limit
theorem which is a statement on the asymptotic distribution of the reduced
variable ${1 \over \sqrt{t}}(\log w_t - \langle \log w_t \rangle)$; 2) the
two necessary and sufficient conditions for $P(w_t)$ to be a power law are
that $\langle  \log \lambda_j \rangle < 0$ (corresponding to a drift $w_t \to
0$) and that $w_t$ not be allowed to become too small. We discuss several models, previously thought unrelated, showing the common underlying mechanism for the generation of power laws by multiplicative processes: the variable $\log w_t$ undergoes a random walk biased to the left but is bounded by a repulsive ''force''.  We give an approximate treatment, which becomes exact for narrow or log-normal distributions of $\lambda$, in terms of the Fokker-Planck equation.  3) For all these models, the exponent $\mu$ is shown exactly to be the solution of $\langle \lambda^{\mu} \rangle = 1$ and is therefore non-universal and
depends on the distribution of $\lambda$. 4) For finite $t$, the power law is
cut-off by a log-normal tail, reflecting the fact that the random walk has
not the time to scatter off the repulsive force to diffusively transport the
information far in the tail.

\vskip 0.5cm

Levy et Solomon ont montr\'e qu'un processus multiplicatif du type  $w_t =
\lambda_1 \lambda_2 ... \lambda_t$  (avec $\lambda_j > 0$) conduit, en
pr\'esence d'une contrainte de bord, \`a une distribution $P(w_t)$ en loi de
puissance $w_t^{-(1+\mu)}$. Nous proposons une d\'erivation simple, intuitive
et exacte de ce r\'esultat bas\'ee sur une analogie avec une marche
al\'eatoire. Nous obtenons les r\'esultats suivants: 1) le r\'egime de loi de
puissance d\'ecrit la distribution asymptotique de $w_t$ aux grands temps et
doit \^etre distingu\'e du th\'eor\`eme limite central d\'ecrivant la
convergence de la variable r\'eduite ${1 \over \sqrt{t}}(\log w_t - \langle
\log w_t \rangle)$ vers la loi gaussienne; 2) les deux conditions
n\'ecessaires et suffisantes pour que $P(w_t)$ soit une loi de puissance sont
$\langle  \log \lambda_j \rangle < 0$ (correspondant \`a une d\'erive vers
z\'ero) et la contrainte que $w_t$ soit emp\'ech\'ee de trop s'approcher de
z\'ero. Cette contrainte peut \^etre mise en oeuvre de mani\`ere vari\'ee,
g\'en\'eralisant \`a une grande classe de mod\`eles le cas d'une barri\`ere
r\'efl\'echissante examin\'e par Levy et Solomon. Nous donnons aussi une
traitement approximatif, devenant exact dans la limite o\`u la distribution de
$\lambda$ est \'etroite ou log-normale en terme d'\'equation de
Fokker-Planck. 3) Pour tous ces mod\`eles, nous obtenons le r\'esultat
g\'en\'eral exact que l'exposant $\mu$ est la solution de l'\'equation
$\langle \lambda^{\mu} \rangle = 1$. $\mu$ est donc non-universel et d\'epend
de la sp\'ecificit\'e de la distribution de $\lambda$. 4) Pour des $t$ finis,
la loi de puissance est tronqu\'ee par une queue log-normale due \`a une
exploration finie de la marche al\'eatoire.

\vskip 0.5cm
Short title: Constrained convergent multiplicative processes
\vskip 0.5cm
PACS:   05.40+j  :  Fluctuations phenomena, random processes and brownian
motion

64.60.Ht : Dynamical critical phenomena

05.70.Ln : Nonequilibrium thermodynamics, irreversible processes

\end{abstract}
\vskip 1cm

\newpage

\section{Introduction}

Many mechanisms can lead to power law distributions.  Power laws have a
special status due to the absence of a characteristic scale and the implicit
(to the physicist) relationship with critical phenomena, a subtle many-body
problem in which self-similarity and power laws emerge from cooperative
effects leading to non-analytic behavior of the partition or characteristic
function.

Recently, Levy and Solomon \cite{Solomon} have presented a novel mechanism
based on random multiplicative processes:
\be
w_{t+1} = \lambda_t w_t  ,
\label{def}
\ee
where $\lambda_t$ is a stochastic variable with probability distribution
$\Pi(\lambda_t)$ and we express $w_t$ in units of a reference value $w_u$
which
could be of the form $e^{rt}$, with $r$ constant. All our analysis below
then
describe the distribution  of $w_t$ normalized to $w_u$, in other words in
the
''reference frame'' moving with $w_u$. At the end, we can easily make
reappear the
scale $w_u$ by replacing everywhere $w$ by ${w \over w_u}$.

Taken litterally with
no other ingredient, expression (\ref{def}) leads to the log-normal
distribution
\cite{GK,Aitcheson,Redner}. Indeed, taking the logarithm of (\ref{def}), we
can
express the distribution of $\log w$ as the convolution of $t$
distributions of
$\log \lambda$. Using the cumulant expansion and going back to the variable
$w_t$
leads, for large times $t$, to
\be
P(w_t)  = {1 \over \sqrt{2\pi Dt}} {1 \over w_t}
\exp \biggl[ - {1 \over 2D t} ( \log w_t  -  v t)^2
\biggl]   ,
\label{lognormalll}
\ee
where $v = \langle \log \lambda \rangle \equiv \int_0^{\infty} d\lambda
\log
\lambda \Pi(\lambda)$ and $D = \langle (\log \lambda)^2 \rangle - \langle
\log
\lambda  \rangle^2$. Expression (\ref{lognormalll}) can be rewritten
\be
P(w_t)  = {1 \over \sqrt{2\pi Dt}} {1 \over w_t^{1 + \mu(w_t)}} e^{\mu(w_t) vt}
\label{approx}
\ee
with
\be
\mu(w_t) = {1 \over 2D t}  \log {w_t \over e^{vt}}  .
\label{muu}
\ee
Since $\mu(w_t)$ is a slowly varying function of $w_t$ this form shows that
the log-normal distribution can be mistaken for an
apparent power law with an exponent $\mu$ slowly varying with the range
$w_t$
which is measured. Indeed, it was pointed out \cite{Montroll} that for $w_t
<<
e^{(v+2D)t}$, $\mu(w_t) << 1$ and the log-normal is undistinguishable from
the
${1 \over w_t}$ distribution, providing a mechanism for $1/f$ noise.
However,
notice that $\mu(w_t) \to \infty$ far in the tail $w_t >> e^{(v+2D)t}$ and
the
log-normal distribution is {\it not} a power law.

The ingredient added by Levy and Solomon \cite{Solomon} is to constrain
$w_t$ to
remain larger than a minimum value $w_0 >0$. This corresponds to put back
$w_t$
to $w_0$ as soon as it would become smaller. To understand intuitively what
happens, it is simpler to think in terms of the variables $x_t = \log w_t$
and
$l = \log \lambda$, here following \cite{Solomon}.
Then obviously, the equation (\ref{def}) defines a random walk in $x$-space
with steps $l$ (positive and negative) distributed according to the density
distribution $\pi(l)=e^l \Pi(e^l)$. The distribution of the position of the
random walk is similarly defined: ${\cal P}(x_t,t) = e^{x_t} P(e^{x_t},t)$.

$\bullet$ If $v \equiv \langle l \rangle = \langle \log \lambda \rangle>
0$, the
random walk is biased and drifts to $+ \infty$. As a consequence, the
presence of
the barrier has no important consequence and we recover the log-normal
distribution
(\ref{lognormalll}) apart from minor and less and less important boundary
effects
at $x_0 = \log w_0$, as $t$ increases. Thus, this regime is without
surprise and
does not lead to any power law. We can however transform this case in the
following one $v \equiv \langle l \rangle < 0$ but a suitable definition of
the
moving reference scale $w_u \sim e^{rt}$ such that, in this frame, the
random
random drifts to the left. But the barrier has to stay fixed in the moving
frame, corresponding to a moving barrier in the unscaled variable $w_t$.

$\bullet$ If $v \equiv \langle l \rangle < 0$, the random walk drifts
towards the
barrier. The qualitative picture is the following (see figures 1 and 2): a
steady-state ($t \to \infty$) establishes itself in which the net drift to
the
left is balanced by the reflection on the reflecting barrier. The random
walk
becomes trapped in an effective cavity of size of order ${D \over v}$ with
an
exponential tail (see below). Its incessant motion back and forth and
repeated
reflections off the barrier and diffusion away from it lead to the build-up
of an
exponential probability (concentration) profile (and no more a gaussian),
as we
are now going to establish, leading in the initial $w_t$ variable to
the announced power law distribution.

\section{The random walk analogy}

In the $x_t = \log w_t$ and $l_t = \log \lambda_t$ variables, expression
(\ref{def}) reads
\be
x_{t+1} = x_t + l_t  ,
\label{defff}
\ee
and describes a random walk with a drift $\langle l \rangle < 0$ to the
left.
The barrier at $x_0 = \log w_0$ ensures that the random walk does not
escape to
$-\infty$. This process is described by the Master equation \cite{Solomon}
\be
{\cal P}(x,t+1) =  \int_{-\infty}^{+\infty}  \pi(l) {\cal P}(x-l,t) dl .
\label{master}
\ee

\subsection{Perturbative analysis}
To get a physical intuition of the underlying mechanism, we now approximate
this
exact Master equation by its corresponding Fokker-Planck equation. Usually,
the
Fokker-Planck equation becomes exact in the limit where the
variance of $\pi(l)$ and the time interval between two steps
go to zero while keeping a constant finite ratio defining the diffusion
coefficient \cite{Risken}. In our case, this corresponds to taking the limit
of very narrow $\pi(l)$ distributions. In this case, we can expand
${\cal P}(x-l,t)$ up to second order
$$
{\cal P}(x-l,t) = {\cal P}(x,t) - l {\partial {\cal P} \over \partial
x}|_{(x,t)} + {1 \over 2} l^2 {\partial^2 {\cal P} \over \partial
x^2}|_{(x,t)}
$$
leading to the Fokker-Planck formution
\be
{\partial {\cal P}(x,t) \over \partial t} = - {\partial j(x,t) \over
\partial x} =
- v {\partial {\cal P}(x,t) \over \partial x} +  D {\partial^2 {\cal
P}(x,t) \over
\partial x^2}  ,
\label{FP}
\ee
where $v = \langle l \rangle$ and $D = \langle l^2 \rangle - \langle l
\rangle^2$ are the leading cumulants of $\Pi(\log \lambda)$.
$j(x,t)$ is the flux defined by
\be
j(x,t) =  v {\cal P}(x,t) -  D {\partial {\cal P}(x,t) \over \partial x} .
\label{flux}
\ee
Expression (\ref{FP}) is nothing but the conservation of probability.
It can be shown that this description
(\ref{FP}) is generic in the limit of very narrow $\pi$ distributions: the
details of $\pi$ are not important for the large $t$ behavior; only its
first two cumulants control the results \cite{Risken}. $v$ and $D$
introduce a characteristic ''length'' ${\hat x} = {D \over |v|}$. In the
overdamped approximation, we can neglect the inertia of the random walker,
and the general Langevin equation $m {d^2x \over dt^2} = -\gamma {dx \over
dt} + F + F_{fluct}$ reduces to
\be
{dx \over dt} = v + \eta(t)  ,
\label{lagngf}
\ee
which is equivalent to the Fokker-Planck equation (\ref{FP})?
$\eta$ is a noise of zero mean and delta correlation with variance
$D$.
This form examplifies the competition between drift $v=- |v|$ and diffusion
$\eta(t)$.

The stationary solution of (\ref{FP}), ${\partial {\cal P}(x,t) \over
\partial t}
= 0$, is immediately found to be
\be
{\cal P}_{\infty}(x) = A - {B \over \mu}  e^{-\mu x} ,
\label{sol}
\ee
with
\be
\mu \equiv { |v| \over D} .
\label{muuytr}
\ee
$A$ and $B$ are two constants of integration. Notice that, as expected in this
approximation scheme, $\mu$ is the inverse of the characteristic length ${\hat
x}$. In absence of the barrier, the solution is obviously $A=B=0$ leading to
the trivial solution $P_{\infty}(x)=0$, which is indeed the limit of
log-normal form (\ref{lognormalll}) when $t \to \infty$. In the presence of
the barrier, there are two equivalent ways to deal with it. The most obvious
one is to impose normalization  \be
\int_{x_0}^{\infty} {\cal P}_{\infty}(x) dx = 1  ,
\label{normefhuf}
\ee
where $x_0 \equiv \log w_0$.
This leads to
\be
{\cal P}_{\infty}(x) = \mu  e^{-\mu (x-x_0)} ,
\label{solutync}
\ee
Alternatively, we can express the condition that the barrier at $x_0$ is
{\it
reflective}, namely that the the flux $j(x_0) = 0$. Let us stress that the
correct boundary condition is indeed of this type (and not absorbing for
instance) as the rule of the multiplicative process is that we put back
$w_t$ to $w_0$ when it becomes smaller than $w_0$, thus ensuring $w_t \geq
w_0$. An absorbing boundary condition would correspond to kill the process
when $w_t \leq w_0$. Reporting (\ref{sol}) in (\ref{flux}) with $j(x_0) =
0$, we retrieve (\ref{solutync}) which is automatically normalized.
Reciprocally,  (\ref{solutync}) obtained from (\ref{normefhuf}) satisfies
the condition  $j(x_0) = 0$.

There is a faster way to get this result (\ref{solutync}) using an analogy
with a
Brownian motion in equilibrium with a thermal bath.
The bias $\langle l \rangle < 0$ corresponds to the existence of a constant
force
$-|v|$ in the $-x$ direction. This force derives from the linearly
increasing
potential $V = |v| x$. In thermodynamic equilibrium, a brownian particle is
found
at the position $x$ with probability given by the Boltzmann factor
$e^{-\beta |v|
x}$. This is exactly (\ref{solutync}) with $D = {1 \over \beta}$ as it
should
from the definition of the random noise modelling the thermal fluctuations.

Translating in the initial variable $w_t = e^x$, we get the Paretian
distribution
\be
P_{\infty}(w_t) = {\mu w_0^{\mu} \over w_t^{1+\mu}} ,
\label{soluui}
\ee
with $\mu$ given by (\ref{muuytr}):
\be
\mu \equiv {|\langle \log \lambda \rangle|
 \over \langle (\log \lambda)^2 \rangle - \langle \log
\lambda  \rangle^2 } .
\label{muuazytr}
\ee

These two derivations should not give the impression that we have found the
exact
solution. As we show below, it turns out that the exponential form is
correct
but the value of $\mu$ given by (\ref{muuazytr}) is only an approximation.
As
already stressed, the Fokker-Planck is valid in the limit of narrow
distributions
of step lengths. The Boltzmann analogy assumes thermal equilibrium, {\it
i.e.}
that the noise is distributed according to a gaussian distribution,
corresponding to a log-normal distribution for the $\lambda$'s. These
restrictive hypothesis are not obeyed in general for arbitrary
$\Pi(\lambda)$.

\subsection{Exact analysis}

In the general case where these approximations do not hold, we have to
address
the general problem defined by the equations (\ref{defff}) and
(\ref{master}).
Let us consider first the case where the barrier is absent. As already
stated,
the random walk eventually escapes to $-\infty$ with probability one.
However,
it will wander around its initial starting point, exploring maybe to the
right and left sides for a while before escaping to $-\infty$. For a given
realization, we
can thus measure the rightmost position $x_{max}$ it ever reached over all
times.
What is the distribution ${\cal P}_{max}(Max(0,x_{max}))$? The question has
been
answered in the mathematical litterature using renewal theory
(\cite{Feller}, p.402) and the answer is
\be
{\cal P}_{max}((Max(0,x_{max})) \sim e^{-\mu x_{max}}  ,
\label{kesth}
\ee
with $\mu$ given by
\be
\int_{-\infty}^{+\infty} \pi(l) e^{\mu l} dl =
\int_0^{+\infty} \Pi(\lambda) \lambda^{\mu} d\lambda = 1  .
\label{cfgtgho}
\ee
 The proof can be sketched in a few lines \cite{Feller} and we summarize it
because it will be useful in the sequel. Consider the   probability
distribution function $M(x) \equiv \int_{-\infty}^x {\cal P}_{max}(x_{max})
dx_{max}$, that  $x_{max} \leq x$. Starting at the origin, this event
$x_{max} \leq x$ occurs if the first step of the random walk verifies $x_1 =
y \leq x$ together with the condition that the rightmost position of the
random walk starting from $-x_1$ is less or equal to $x-y$. Summing over all
possible $y$, we get the Wiener-Hopf integral equation
\be
M(x) = \int_{-\infty}^x M(x-y) \pi(y) dy  .
\label{hofgt}
\ee
It is straightforward to check that $M(x) \to e^{-\mu x}$ for large $x$ with
$\mu$ given by (\ref{cfgtgho}). We refer to \cite{Feller} for the
questions of uniqueness and to \cite{Morse,Frisch} for classical methods for
handling Wiener-Hopf integral equations. We shall encounter the same type of
Wiener-Hopf integral equation in section 3.3 below which addresses the
general case.

How is this result useful for our problem? Intuitively, the presence of the
barrier, which prevents the escape of the random walk, amounts to reinjecting the
random walker and enabling it to sample again and again the large positive
deviations described by the distribution (\ref{kesth}). Indeed, for such a large
deviation, the presence of the barrier is not felt and the presence of the drift
ensures the validity of (\ref{kesth}) for large $x$. These intuitive
arguments are
shown to be exact in section 3.3 for a broad class of processes.

Let us briefly mention that there is another way to use this problem, on the
rightmost  position $x_{max}$ ever reached, to get an exponential distribution
and  therefore a power law distribution in the $w_t$ variable. Suppose that we
have a constant input of random walkers, say at the origin. They establish a
uniform flux directed towards $-\infty$. The density (number per unit length) of
these walkers to the right is obviously decaying as given by (\ref{kesth}) with
(\ref{cfgtgho}). This provides an alternative mechanism for generating
power laws,
based on the superposition of many convergent multiplicative processes.

Let us now compare the two results (\ref{muuazytr}) and (\ref{cfgtgho}) for
$\mu$.
It is straightforward to check that (\ref{muuazytr}) is the solution of
(\ref{cfgtgho}) when $\pi(l)$ is a gaussian {\it i.e.}  $\Pi(\lambda)$ is a
log-normal distribution. (\ref{muuazytr}) can also be  obtained perturbatively
from (\ref{cfgtgho}): expanding $e^{\mu l}$ as $e^{\mu l} = 1 + \mu l + {1 \over
2} \mu^2 l^2 + ...$ up to second order, we find that the solution of
(\ref{cfgtgho}) is (\ref{muuazytr}). This was expected from our previous
discussion of the approximation involved in the use of the Fokker-Planck
equation.

\subsection{Relation with Kesten variables}

Consider the following mixture of multiplicative and additive process:
\be
S_{t+1}
= b_t + \lambda_t S_t  ,       \label{jekentin}
\ee
with $\lambda$ and $b$ are positive independent random variables. The
stochastic
dynamical process (\ref{jekentin}) has been introduced in
various occasions, for instance in the physical modelling of 1D
disordered systems \cite{Calan} and the statistical representation of
financial time series \cite{Haan}. The variable $S(t)$ is known in
probability
theory as a Kesten variable \cite{Kesten}.

Notice that $b=0$ recovers (\ref{def}). It is well-known that for $\langle
\log
\lambda \rangle < 0$, $S(t)$ is distributed according to
a power law  \be
P(S_t) \sim S_t^{-(1+\mu)}  ,
\label{pofhgu}
\ee
with $\mu$ determined by the condition (\ref{cfgtgho}) \cite{Kesten}
already encountered above $\langle \lambda^{\mu} \rangle = 1$. In fact, the
derivation of (\ref{pofhgu}) with (\ref{cfgtgho}) uses the result
(\ref{kesth})
of the renewal theory of large positive excursions of a random walk biased
towards $-\infty$ \cite{Haan}. The figure 3 shows the reconstructed
probability density of the kesten variable $S_t$ for $\lambda_t$ and
$b_t$ uniformely sampled in the interval $[0.48;1.48]$ and in $[0,1]$
respectively. This corresponds to the theoretical value $\mu \approx 1.47$.
We have also constructed the
probability density function of the variations $S_{t+1} - S_t$ of the kesten
variable for the same values. We observe again a powerlaw tail for
the positive and negative variations, with the same exponent .

This is not by chance and we now show that the multiplicative process with the
reflective barrier and the Kesten variable are deeply related. First,
notice that
for  $\langle \log \lambda \rangle < 0$ in absence of $b(t)$, $S_t$ would
shrink to
zero. The term $b(t)$ can be thought of as an effective repulsion from zero and
thus acts similarly to the previous barrier $w_0$. To see this more
quantitatively,
we form
\be
{S_{t+1} - S_t \over S_t} = {b_t \over S_t} + \lambda_t - 1  ,
\label{jekentrtin}
\ee

We make the approximation of writing the finite difference ${S_{t+1} - S_t \over
S_t}$ as ${d \log S \over dt}$. It has the same status as the one used to
derive
the Fokker-Planck equation and will lead to results correct up to the
second cumulant. Introducing again the variable $x \equiv \log S$, expression
(\ref{jekentrtin})
gives the overdamped Langevin equation:
\be
{dx \over dt} =   b(t) e^{-x} - |v| + \eta(t)  ,
\label{lagnsfgf}
\ee
where we have written $\lambda(t) - 1$ as the sum of its mean and a purely
fluctuating part. We thus get $v = \langle \lambda \rangle - 1 \simeq
\langle \log \lambda \rangle$ and
$D \equiv \langle \eta^2 \rangle = \langle \lambda^2 \rangle -
\langle \lambda \rangle^2 \simeq \langle \log (\lambda)^2 \rangle -
\langle \log \lambda \rangle^2$.
Compared to (\ref{lagngf}), we see the additional term $b(t) e^{-x}$,
corresponding to a repulsion from the $x<0$ region. This repulsion
replaces the reflective barrier, which can itself in turn be
modelled by a concentrated force. The corresponding
Fokker-Planck equation is
\be
{\partial P(x,t) \over \partial t} = b(t) e^{-x} P(x,t)
- (v + b(t)e^{-x})  {\partial P(x,t) \over \partial x} +  D {\partial^2
P(x,t)
\over \partial x^2}  .
 \label{FrP}
\ee
It also presents a well-defined stationary solution that we can easily
obtain in
the regions $x \to +\infty$ and $x \to - \infty$. In the first case,
the terms $b(t) e^{-x}$ can be neglected and we recover the previous
results
(\ref{solutync}) with $x_0$ now determined from asymptotic matching with
the
solution at $x \to -\infty$. For $x \to -\infty$, we can drop all the terms
except those in factor of the exponentials which diverge and get
$P(x) \to e^{x}$. Back in the $w_t$ variable, $P_{\infty}(S_t)$ is a
constant for
$S_t \to 0$ and decays algebraically as given by (\ref{soluui}) with the
exponent (\ref{muuytr},\ref{muuazytr}) for $S_t \to + \infty$.

Beyond these approximations, we can solve exactly expression
(\ref{jekentrtin}) or
equivalently (\ref{jekentin}) and we recover (\ref{cfgtgho}). This is
presented
in section 3.3 below. Again, notice that
(\ref{muuytr},\ref{muuazytr}) is equal to the solution of (\ref{cfgtgho})
up to
second order in the cumulant expansion of the distribution of $\log
\lambda$.

It is interesting to note that the Kesten process (\ref{jekentin}) is a
generalization of branching processes \cite{Harris}. Consider the simplest
example of a branching process in which a branch can either die with
probability
$p_0$ or give two branches with probability $p_2 = 1 - p_0$. Suppose in
addition that, at
each time step, a new branch nucleates. Then, the number of branches
$S_{t+1}$ at
generation $t+1$ is given by eq.(\ref{jekentin}) with $b_t=1$ and
$\lambda_t =
{2j_{t+1} \over S_t}$, where $j_{t+1}$ is the number of branches out of the
$S_t$
which give two branches. The distribution $\Pi(\lambda)$ is simply deduced
from
the binomial distribution of $j_{t+1}$, namely ${S_t \choose j_{t+1}}
p_0^{j_{t+1}} p_2^{S_t-j_{t+1}}
\equiv
{[S_t]! \over [S_t-j_{t+1}]! [j_{t+1}]!} p_0^{j_{t+1}} p_2^{S_t-j_{t+1}}$.
For large $S_t$, $\Pi(\lambda)$ is
approximately a gaussian with a standard deviation  equal to ${4p_0(1-p_0)
\over
S_t}$, {\it i.e.} it goes to zero for large $S_t$. We thus pinpoint here
the key
difference between standard branching processes and the Kesten model: in
branching models, large generations are {\it self-averaging} in the sense
that the
number of children at a given generation fluctuates less and less as the
size of
the generation increases, in contrast to eq.(\ref{jekentin})
exhibiting  the same {\it relative} fluctuation amplitude. This is the
fundamental reason for
the robustness of the existence of a power law distribution in contrast to
branching models in which a power law is found only for the special
critical case
$p_0 = p_2$ at the edge of the run away condition. The same conclusion
carries out
directly for more general branching models. Note finally that it can be
shown that the branching
model previously defined becomes equivalent to a Kesten process if the
number of branches formed from a single one is itself a random variable
distributed according to a power law with the special exponent $\mu=1$,
ensuring
the scaling of the fluctuations with the size of the generations.

\subsection{Generalization to a broad class of multiplicative process with
repulsion at the origin}

The above considerations lead us to propose the following generalization
\be
w_{t+1} = e^{f(w_t, \{\lambda_t, b_t,...\})} \lambda_t w_t  ,
\label{general}
\ee
where $f(w_t, \{\lambda_t, b_t,...\})) \to 0$ for $w_t \to \infty$ and
$f(w_t, \{\lambda_t,b_t,...\})) \to \infty$ for $w_t \to 0$.

The model (\ref{def}) is the special case $f(w_t, \{\lambda_t, b_t,...\}) = 0$
for $w_t > w_0$ and $f(w_t, \{\lambda_t, b_t,...\}) = \log ( {w_0 \over
\lambda_t
w_t}) $ for $w_t \leq w_0$. The Kesten model (\ref{jekentin}) is the
special case
$f(w_t, \{\lambda_t, b_t,...\}) = \log (1 + {b(t) \over \lambda_t w_t})$. More
generally, we can consider a process in which at each time step $t$, after the
variable $\lambda_t$ is generated, the new value $ \lambda_t w_t$ (or $
\lambda_t
w_t + b_t$ in the case of Kesten variables) is readjusted by a factor
$e^{f(w_t,\{\lambda_t, b_t,...\})}$  reflecting the constraints imposed on the
dynamical process. It is
thus reasonable to consider the case where $f(w_t, \{\lambda_t, b_t,...\})$
depends on $t$ only through the dynamical variables $\lambda_t$ (and in special
cases $b_t$), a condition which already holds for the two examples above. In the
following Fokker-Planck approximation, we shall consider the case where  $f(w_t,
\{\lambda_t, b_t,...\})$ is actually a function of the product $ \lambda_t w_t$,
which is the value generated by the process at step $t$ and to which the
constraint
represented by $f(\lambda_t w_t)$ is  applied. We shall turn back to the
general case (\ref{general}) in section 2.5.

In the Fokker-Planck approximation, $f(\lambda_t w_t)$ defines an effective
repulsive
stochastic force. To illustrate the repulsive mechanism, it is enough to
consider the restricted case where $f(w_t)$ is only a function of $w_t$.
This corresponds to freezing the random part in the noise term $\lambda_t$
leading
to the definition of the diffusion coefficient. In the random walk analogy, we
thus have
the force $F(x_t) = f(w_t)$ acting on the random walker. The corresponding
Fokker-Planck equation is
\be
{\partial {\cal P}(x,t) \over \partial t} =
-  {\partial (v + F(x)) {\cal P}(x,t) \over \partial x} +  D {\partial^2
{\cal
P}(x,t) \over \partial x^2}  .
\label{FddP}
\ee
$F(x)$ decays to zero at $x \to \infty$ and establishes a repulsion
of the diffusive process in the negative $x$ region: this is the
translation in
the random walk analogy of the condition $f(w_t) \to
\infty$ for $w_t \to 0$.

With these properties, the tail of ${\cal P}(x)$ for large $x$ and large
times is
given by  ${\cal P}_{\infty}(x) \sim  e^{-\mu x}$, and as a consequence
$w_t$ is
distributed  according to a power law, with exponent $\mu$ given again
approximately by (\ref{muuytr},\ref{muuazytr}). The shape of the potential
defined
by $v + F(x) = -{\partial V(x) \over \partial x}$, showing the fundamental
mechanism, is depicted in figure 4. As we have already noted, the bound $w_0$
leading to a reflecting barrier is a special case of this general situation,
corresponding to a concentrated repulsive force at $x_0$.

The expression (\ref{general}) for the general model can be ''derived'' from
the overdamped Langevin equation equivalent to the Fokker-Planck equation
(\ref{FddP}):
 \be
{dx \over dt} = F(x) - |v| + \eta(t)  .
\label{ltngf}
\ee
Let us take the discrete version of (\ref{ltngf}) as $x_{t+1} = x_t +
F(x_t)  -
|v| + \eta_t$, replace with $x_t = \log w_t$ and exponentiate to obtain
\be
w_{t+1} = e^{F(\log w_t)} \lambda_t w_t  ,
\label{modeljf}
\ee
where $\lambda_t \equiv e^{- |v| + \eta_t}$. Since $F(x) \to 0$ for large $w_t$,
we recover a pure multiplicative model $w_{t+1} = \lambda_t w_t$ for the
tail. The
condition that $F(x)$ becomes very large for negative $x$ ensures that $w_t$
cannot decrease to zero as it gets multiplied by a diverging number when it goes
to zero.

\subsection{Exact derivation of the tail of the power law distribution}

The existence of a limiting distribution for $w_t$ obeying
(\ref{general}), for a large class of $f(w, \{\lambda, b,...\})$ decaying
to zero for large $w$ and going to infinity for $w \to 0$, is ensured by the
competition between the convergence of $w$ to zero and the sharp repulsion
from it. We shall also suppose in what follows that $\partial f(w, \{\lambda, b,...\})/\partial x \to 0$ for $w \to \infty$, which is satisfied for a large class of smooth functions already satisfying the above conditions.
It is an interesting mathematical problem to establish this result rigorously,
for instance by the method used in \cite{Solomon,Frisch}.
Assuming the existence of the asymptotic distribution $P(w)$, we can
determine its
shape, which must obey
\be
v \equiv w e^{-f(w, \{\lambda, b,...\})} \eql \lambda w  ,
\label{laddfl}
\ee
where $\{\lambda, b,...\}$ represents the set of stochastic variables used
to define the random process. The expression (\ref{laddfl}) means that
the l.h.s. and r.h.s. have the same distribution. We can thus write
$$
P_v(v) = \int_0^{+\infty} d\lambda \Pi(\lambda) \int_0^{+\infty}  dw P_w(w)
\delta(v-\lambda w) =
\int_0^{+\infty} {d\lambda \over \lambda} \Pi(\lambda) P_w({v \over \lambda}) .
$$
Introducing $V = \log v$, $x \equiv \log w$ and $l \equiv \log \lambda$, we
get
\be
P(V) = \int_{-\infty}^{+\infty} dl \Pi(l)  P_x(V-l)  ,
\label{tresifim}
\ee

Taking the logarithm of (\ref{laddfl}), we have
$V = x - f(x, \{\lambda, b,...\})$, showing that $V \to x$ for large $x>0$,
since
we have assumed that $f(x, \{\lambda, b,...\}) \to 0$ for large $x$. We can
write
$P(V) dV = P_x(x) dx$ leading to $P(V) =
{ P_x(x(V)) \over {1 - \partial f(x, \{\lambda, b,...\})/\partial x}} \to  P_x(V)$ for
$x \to \infty$. We thus recover the Wiener-Hopf integral equation (\ref{hofgt})
yielding the announced results (\ref{kesth}) with (\ref{cfgtgho}) and therefore
the power law distribution (\ref{soluui}) for $w_t$ with $\mu$ given by
(\ref{cfgtgho}).

This derivation explains the origin of the generality of these results to
a large class of convergent multiplicative processes repelled from the
origin.

\section{Discussion}

\subsection{Nature of the solution}

To sum up, convergent multiplicative processes repelled from the origin
lead to power law distributions for the multiplicative variable $w_t$
itself.
Ideally, this holds true in the asymptotic regime, namely after an infinite
number of stochastic products have been taken. This addresses a different
question than that answered by the log-normal distribution for
unconstrainted
processes which describes the convergence of the reduced
variable ${1
\over \sqrt{t}} (\log w_t -\langle \log w_t \rangle)$ to the gaussian law.
Notice that this reduced variable tends  to zero for our problem
and thus does not contain any useful information.

\subsection{The exponent $\mu$}

In the Fokker-Planck approximation of the random walk analogy, $\mu$ is the
inverse of the size of the effective cavity trapping the random walk. In
this
approximation, $\mu$ is a function of, and
only of, the first two cumulants of the distribution of $\log \lambda$. In
particular, if the drift $|v| < 2D$, $\mu < 2$ corresponding to variables
with no
variance and even no mean when $\mu < 1$ ($|v| < D$). It is rather
intuitive:
large fluctuations in $\lambda$ leads to a large diffusion coefficient $D$
and
thus to large fluctuations in $w_t$ quantified by a small $\mu$. Recall
that the smaller $\mu$ is, the wilder are the fluctuations.

Within an exact formulation, we have shown that there is a rather subtle
phenomenon which identifies $\mu$ as the inverse of the typical value of the largest excursion against the flow of a  particle in random motion with drift.  This holds true
for a
large class of models  characterized by a negative drift and a
sufficiently
fast repulsion from the negative domain (in the $x$-variable), {\it i.e.}
from the
origin (in the $w$-variable).

\subsection{Dynamical interpretation}

There are two ways to interpret the power law distribution we obtain. The
first
one is to generate an ensemble of multiplicative processes and, in
practice, examine their values after a large number of multiplications. The
power law then
describes the distribution of the set of values thus obtained. The
second
interpretation is to consider the time evolution of a single constrained
multiplicative process and examine the distribution of the values it takes
over time. In the stationary regime these two distributions coincide. The
models studied here correspond to
 the latter case. By the
law
of convolution, it is straightforward to check that the distribution of
variations over one or more time steps of the variables is also a power law
with
the same exponent.

At large $t$, $w_t$ thus does not converge but goes on
fluctuating as shown in figure 2. What does converge is its distribution
$P(w_t)
\to P_{\infty}(w_t)$ given by (\ref{soluui}). $w_t$ is a dynamical
variable
distributed with a power law. As can be seen from eq.(\ref{soluui}), the
{\it
typical} scale is set by $w_0$ for the reflecting barrier problem and by 
$\langle b
\rangle$ for the Kesten variable. It is however clear from inspection of
figure 2b that most of the time $w_t$ is less than its average. It also shows
rare intermittent excursions to much larger values.

\subsection{Additional constraint fixing $\mu$}

We recover the relationship
relating $\mu$ to the minimum value $w_0$ in the reflecing barrier problem
by
specifying \cite{Solomon} the value $C$ of the average  $\langle w_t
\rangle$.
Calculating the average straightforwardly using (\ref{soluui}), we get
$\langle
w_t \rangle = w_0 {\mu \over \mu - 1}$, leading to
\be
\mu = {1 \over 1 - {w_0 \over C}} .
\ee
Notice that this expression is a special case of (\ref{cfgtgho}) and should
by
no mean be interpreted as implying that $\mu$ is controlled by $w_0$ in
general.
This is only true with an {\it additional} constraint, here of fixing the
average.
The {\it general} result is that $\mu$ is given by (\ref{cfgtgho}), {\it
i.e.}
at a minimum by the two first cumulents of the distribution of $\log
\lambda$.

\subsection{Positive drift in the presence of an upper bound}

Consider a purely multiplicative process
where the drift is reversed $\langle \log \lambda \rangle > 0$,
corresponding to
an average exponential growth of $w_t$ in the presence of a barrier $w_0$
limiting
$w_t$ to be {\it smaller} than it. The same reasoning holds and a parallel
derivation yields \be
P_{\infty}(w_t) = {\mu \over w_0^{\mu}} w_t^{\mu - 1} ,
\label{soluasdwui}
\ee
with $\mu \geq 0$ again given by (\ref{cfgtgho}). This distribution
describes the
values $0 < w_t < w_0$. Notice that, if $\mu > 1$, the distribution is
{\it increasing} with $w_t$. This is obviously no more a power law of the
tail, rather a power law for the values close to zero. For $\mu < 1$,
$P_{\infty}(w_t)$ decays as a power law, however bounded by $w_0$ and
diverging at zero (while remaining safely normalized). This shows that,
when
speaking of general power law distribution for large values, this regime is
not
relevant. Only the regime with negative drift and lower bound is
relevant.\\

However, in the case of Kesten variables (\ref{jekentrtin}), if $S_t$ is
growing exponentially with an
average rate $\langle \log \lambda_t \rangle > 0$, and if the input flow
$b_t$ is
also increasing with a {\it larger} rate $r$, we
define $b_t = e^{r(t+1)} {\hat b}_t$, where ${\hat b}_t$ is a stochastic
variable of order one. We also define $\lambda_t = {\hat \lambda}_t e^{r}$. If
$r > \langle \log a_t \rangle$, then $\langle \log {\hat \lambda}_t \rangle
< 0$.

The equation (1) thus transforms into ${\hat S}_{t+1} = {\hat \lambda}_t {\hat
S}_t +
{\hat b}_t$, with $S_t = e^{rt}  {\hat S}_t$, and where ${\hat \lambda}_t $ and
${\hat b}_t$ obey exactly the conditions for our previous analysis to
apply. The
conclusion is that, due to input growing exponentially fast, the growth
rate of $w_t$ becomes that of the input, its average
(which exists for $\mu > 1$) grows exponentially as
$\langle S_t \rangle \sim e^{rt}$ and its value exhibits large fluctuations
governed by the power law pdf $P(S_t) \sim {e^{\mu rt} \over S_t^{1+\mu}}$
with $\mu$ solution of $\langle \lambda_t^{\mu} \rangle = e^{r\mu}$, leading
to $\mu = {\langle b_t \rangle - \langle \lambda_t \rangle \over
\langle \lambda_t^2 \rangle - \langle \lambda_t \rangle^2}$ in the second
order cumulant approximation.

\subsection{Transient behavior}

For $t$ large but finite, the exponential (\ref{tdfbc}) is
truncated  and decays typically like a gaussian for $x > \sqrt{Dt}$.
Translated in
the $w_t$ variable, the power law distribution (\ref{soluui}) extends up to
$w_t \sim e^{\sqrt{Dt}}$ and transforms into an approximately log-normal
law for large values. Refining these results for finite $t$ using the theory of
renewal processes is an interesting mathematical problem left for the future.

\subsection{Non-stationary processes}

 When the multiplicative process
(\ref{def}) is not stationary in time, for instance if $v(t), D(t)$ or
$x_0(t)$ become function of time, then their characteristic time $\tau$ of
evolution must be compared with $t^*(x) = {x^2 \over D}$. For ''small'' $x$ such
that $t^*(x) << \tau$, the distribution $P(x,t)$ keeps an exponential tail
with an
exponent adiabatically following  $v(t), D(t)$ or $x_0(t)$. We thus predict
a power
law distribution for $w_t$ but with an exponent varying with $v$ and $D$
according to eq.(\ref{muuytr},\ref{muuazytr}).
For ''large'' $x$ such that $t^*(x) \geq \tau$, the diffusion process has
not time to reach $x$ and to bounce off the barrier that the
parameters have alreay changed. It is important to stress again the physical
phenomenon at the origin of the establishement of the exponential profile: the
repeated encounters of the diffusing particle with the barrier. For large
$x$, the
repeated encounters take a large time, the time to diffuse from $x$ to the
barrier
back and forth. In this regime $t^*(x) \geq \tau$, the exponential profile
for $P(x)$ has not time to establish itself since the parameters of the
diffusion evolve faster that the ''scattering time'' off the barrier.
The analysis of the modification of the tail in the presence of
non-stationarity effects is left to a separate work. In particular, we
would like
to understand what are the processes which lead to an exponential cut-off of the
power law in the $w_t$ variable, corresponding to an exponential of an
exponential
cut-off in the $x$-variable.

\subsection{Status of the problem}

 Levy and Solomon \cite{Solomon} propose that
the power law (\ref{soluui}) is to multiplicative processes what the
Boltzmann distribution is to additive processes. In the latter case, the
fluctuations can be described by a single parameter, the temperature
($\beta^{-1}$) defined from the factor in the Boltzmann distribution
$e^{-\beta E}$. In a nutshell,
recall that the exponential Boltzmann distribution stems from the fact that
the number $\Omega$ of microstates constituting a macro-state in an
equilibrium system is multiplicative in the number of degrees of freedom while
the energy $E$ is additive. This holds true when a system can be partitionned
into weakly interactive sub-systems. The only solution of the resulting
functional equation $\Omega(E_1 +
E_2) = \Omega(E_1) \Omega(E_2)$ is the exponential.

No such principle applies in
the multiplicative case. Furthermore, the Boltzmann reasoning that we have
used
in section 2.1 is valid only under restrictive hypotheses and provides at
best an
approximation for the general case. We have shown that the correct exponent
$\mu$
is in fact controlled by {\it extreme} excursions of the drifting random
walk
against the main ''flow'' and not by its average behavior. This rules out
the
analogy proposed by Levy and Solomon.

\section{Conclusion}

We believe we have clarified and generalized the mechanism found by  Levy and
Solomon \cite{Solomon} to obtain power law distributions from  constrained
multiplicative process. The basic mechanism resides in the competition
between an
average drift to zero compensated by a mechanism tending to restore the
multiplicative variable to a finite value. This reminds us of the interplay
between exponential sensitivity to initial conditions and nonlinear
reinjection at
the heart of chaotic behavior. We have presented an intuitive
approximate derivation of the power law distribution and its exponent, using the
Fokker-Planck formulation in a random walk analogy. Our main
result is the explicit calculation of the exponent of the power law
distribution,
as a solution of a Wiener-Hopf integral equation, showing that it is
controlled by
extreme values of the process. For narrow distributions of the factors $\lambda$'s, $\mu$
reduces approximately to the ratio of its first two cumulants. We have also been
able to extend the initial problem to a large class of systems where the common
feature is the existence of a mechanism repelling the variable away from
zero. We
have in particular drawn a connection with the Kesten process well-known to
produce power law distributions. The results presented in this paper are of
importance for the description of many systems in Nature showing complex
intermittent self-similar dynamics. This will be addresses in a separate
communication.

\vskip 0.5cm
{\bf Acknowledgments}
\vskip 0.5cm
D.S. acknowledges stimulating correspondence with S. Solomon and U. Frisch.
This work is Publication no. 4642 of the Institute
of Geophysics and Planetary Physics, University of California, Los Angeles.

\pagebreak

FIGURE CAPTIONS :
\vskip 1cm

Fig.1 : Steady-state exponential profile of the probability density of
presence of
the random walk with a negative drift and a reflecting barrier.

\vskip 0.5cm

Fig.2 : a) A typical trajectory of the random walker at large times, showing the
multiple reflections off the barrier. b) The time evolution of the Kesten
variable
defined by the equation (\ref{jekentin}) with $a_t$ uniformely taken in the
interval $[0.48;1.48]$ leading to $\mu \approx 1.47$ according to
(\ref{cfgtgho})
and $b_t$ uniformely taken in the interval $[0;1]$. Notice the intermittent
large
excursions.

\vskip 0.5cm
Fig.3 : Reconstructed natural logarithm of the
probability density of the kesten variable $S_t$ as a function of the
natural logarithm of $S_t$, for $0.48 \leq \lambda_t \leq 1.48$ and $0 \leq
b_t \leq 1$, uniformely sampled. The theoretical prediction $\mu \approx 1.47$
from (\ref{cfgtgho}) is quantitatively verified.

\vskip 0.5cm
Fig.4 : Generic form of the potential whose gradient gives the force felt by the
random walker. This leads to a steady-state exponential profile of its density
probability, corresponding to a power law distribution of the $w_t$-variable.


\begin{thebibliography}{99}

\bibitem{Solomon} M. Levy and S. Solomon, Power laws are logarithmic Boltzmann
laws, to appear in Int. J. Mod. Phys. C; Of wealth power and law: the origin
of scaling in Economics, submitted to Nature.

\bibitem{GK} B.V. Gnedenko and Kolmogorov, A.N. {\em Limit distributions for
sum of independent random variables}, Addison Wesley, Reading MA (1954).

\bibitem{Aitcheson} J. Aitcheson and J.A.C. Brown, The log-normal distribution
(Cambridge University Press, London, England, 1957).

 \bibitem{Redner} S. Redner, Random multiplicative processes:
An elementary tutorial, Am. J. Phys. 58, 267-273, 1990.

\bibitem{Montroll} E.W. Montroll and M.F. Shlesinger, Proc. Nat. Acad. Sci.
USA 79, 3380-3383 (1982).

\bibitem{Risken}  H. Risken, The Fokker-Planck equation: methods of solution
and applications, 2nd ed. (Berlin; New York : Springer-Verlag, 1989).

\bibitem{Feller} W. Feller, An introduction to probability theory and its
applications, Vol II, second edition (John Wiley and sons, New York, 1971).

\bibitem{Morse} P.M. Morse and H. Feshbach, Methods of theoretical physics
(McGraw Hill, New York, 1953).

\bibitem{Frisch} H. Frisch, J. Quant. Spectrosc. Radiat. Transfer 39, 149
(1988).

\bibitem{Calan} C. de Calan, J.-M. Luck, Th. M. Nieuwenhuizen and D. Petritis,
On the distribution of a random variable occurring in 1D disordered systems,
J.Phys.A 18, 501-523, 1985.

\bibitem{Haan} L. de Haan, S.I. Resnick, H. Rootz\'en and C.G. de Vries,
Extremal behavior of solutions to a stochastic difference equation with
applications to ARCH processes, Stochastic Processes and their Applications
32, 213-224, 1989.

\bibitem{Kesten} H. Kesten, Random difference equations and renewal theory for
products of random matrices, Acta Math. 131, 207-248, 1973.

\bibitem{Harris} T.E. Harris, The theory of branching processes (Springer,
Berlin,
1963).



\end{thebibliography}
\end{document}